\documentclass[aps,prb,twocolumn,epsf,floatfix]{revtex4}
\usepackage{graphicx}
\usepackage{epsfig}
\begin{document}

\title{Robustness of Decoherence-Free States for Charge Qubits 
under Local Non-uniformity}
\author{Tetsufumi Tanamoto and Shinobu Fujita}
%} 
\affiliation{Advanced LSI Technology Laboratory, %Corporate R\&D Center, 
Toshiba Corporation,
Saiwai-ku, Kawasaki 212-8582, Japan}
%\draft

\begin{abstract}
We analyze the robustness of decoherence-free (DF) subspace and subsystem 
in charge qubits, when difference from the collective decoherence measurement
condition is large ($\sim$5\%) 
in the long time period, which is applicable 
for solid-state qubits using as a quantum memory. 
We solve master equations of up to four charge qubits and a detector
as a quantum point contact (QPC). We show that 
robustness of DF states is strongly affected by local non-uniformities. 
We also discuss the possible two-qubit logical states by exactly 
solving the master equations. 
\end{abstract}
\maketitle

%%%%%%%%%%%%%%%%%%%%%%%%%%%%%%%%%%%%%%%%%%%%%%%%%%%%%
%%%%%%%%%%%%%%%%%%%%%%%%%%%%%%%%%%%%%%%%%%%%%%%%%%%%%
%{\it Introduction}--
Although decoherence is the largest obstacle for quantum 
information processing, a lot of powerful active 
methods  for correcting effects of 
decoherence have been discovered\cite{Chuang}.
As to the passive anti-decoherence protection, 
{\it decoherence-free} (DF) {\it subspace}\cite{Zanardi,Lidar} 
and {\it subsystem}\cite{Knill} are shown to be  very useful for 
{\it collective decoherence environment}, in which 
all qubits suffer the same disturbance from the environment.
Singlet state is the only DF state for two qubits and 
there are two independent DF subspace bases for four qubits, {\it e.g.}
\begin{eqnarray}
|\Psi_1^{[4]}\rangle_{(1234)}\! &\!=\!&\!
 2^{-1}(|01\rangle\!-\!|10\rangle)_{(12)}
 \otimes (|01\rangle\!-\!|10\rangle)_{(34)},\ \ \ \ \nonumber \\
|\Psi_2^{[4]}\rangle_{(1234)}&=&1/(2\sqrt{3})
(2|0011\rangle-|0101\rangle-|0110\rangle-
|1001\rangle \nonumber \\
&-&|1010\rangle+2|1100\rangle)_{(1234)}
\label{wavefun}
\end{eqnarray}
where $|1001\rangle_{(1234)}\!=\!
|1\rangle_{1}|0\rangle_{2}|0\rangle_{3}|1\rangle_{4}$ and so on.
The DF subsystem starts from three qubits.
Experiments have succeeded until four qubits in 
photon system\cite{Bourennane,Altepeter} and three qubits in
nuclear magnetic resonance (NMR)\cite{Viola}.
Bacon{\it et al.}\cite{Lidar} also showed that, 
even if there is a symmetry breaking perturbation from the collective 
environment, which is parameterized by a coupling strength $\eta$, 
the DF subspace is robust in the order of $O(\eta)$ when $\eta \!\ll\!1$.

However, in the case of solid-state qubits, even a single qubit 
is hard to fabricate and the redundancy regarding the number 
of qubits would be a critical issue in constructing a large qubit system.
First of all, we could not prepare plenty of qubits with 
mathematically exact size. The sizes of Cooper-pair box of Ref.\cite{Nakamura}
and GaAs quantum dot (QD)\cite{Hayashi} where coherent oscillation 
can be observed are less than hundreds of nm. The requirement of 
a few \% fluctuation between qubits would result in controllability of
a few nm in fabrication. This would be unrealistic 
until future when fabrication process is greatly 
advanced. 
The fluctuation of sizes would lead to that of interaction 
amplitude between qubits and a measurement apparatus, and 
that of the applied gate bias, 
in addition to the effect of randomly distributed background traps
\cite{Makhlin,You}.
Note that even a roughness of the order of 1 \AA \ at interfaces 
affects current characteristics in advanced LSI technologies 
as shown in Ref.\cite{Uchida}
Thus, the non-uniformity of solid-state qubits to collective decoherence environment is much larger and rather {\it local} compared 
with that of optical or NMR qubits, 
and it will be necessary to consider the effects of second order  
$O(\eta^2)$ or higher symmetry-breaking perturbation. 
%%%%%%%%%%%%%%%%%%%%%%%%%%%%%%%%%%%%%%%%%%%%%%%%%%%%%%%%%%%%%%%%%
%%%%%%%%%%%%% Fig.1
\begin{figure}
\begin{center}
\includegraphics[width=7.5cm]{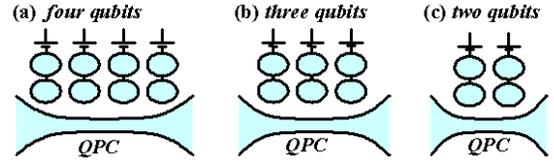}
\caption{Qubits that use double dot charged states are capacitively 
coupled to a QPC detector. 
}
\end{center}
\label{QPC}
\end{figure}
%%%%%%%%%%%%%%%%%%%%%%%%%%%%%%%%%%%%%%%%%%%%%%%%%%%%%%%%%%%%%%%%%

In this paper, we theoretically describe the effect of 
large non-uniformity ($\sim$5\%) 
of charge qubit parameters on DF states  
based on time-dependent density matrix (DM) equations considering the 
measurement process by detector current. 
Detection of qubit states induces a {\it backaction} on the qubit state 
resulting in a corrupting of qubit states. Thus, the measurement is an 
important interaction with the environment for qubits.
The charge qubit is a two-level system\cite{Landauer} controlled by
gate electrodes\cite{Makhlin,Tanamoto}, and constituted from 
coupled QDs where one excess electron is inserted, 
assuming there is one energy level in each QD.
The detector discussed here is a quantum point contact (QPC) 
in the tunneling region depicted in Fig.1.
The position of the excess charge affects the QPC current electrically, 
resulting in detection of charged state. 
%- This would be one of the common setups in quantum circuits.
Experiments\cite{Field,Vandersypen} have successfully proved the 
high sensitivity of QPC detector current
%- Field {\it et al}\cite{Field} 
%- and Gardelis {\it et al.}\cite{Gardelis} succeeded in 
%- measurement of charging of a QD by capacitively coupled detector current 
$I=G_0Tv_d$ ($G_0=2e^2/h\sim 77 \mu$S, 
$T$ is a transmission coefficient $v_d$ is a bias between electrodes)
in the tunneling region($T<1$).
The QPC current induces {\it shot noise} as 
the basic and unavoidable noise\cite{Vandersypen}, 
and the cause of decoherence treated here. 
%- We estimate energy unit by $\Gamma_0\equiv I/e=G_0v_d/e\sim4.836\times10^9$Hz.
%%%%%%%%%%%%%%%%%%%%%%%%%%%%%%%
%%%%%%%%%%%%%%%%%%%%%%%%%%%%%%%
The purpose of this study is to investigate 
the robustness of many-qubit DF states for 
{\it local} large non-uniformity and show the possibility of 
using non-DF states with two-qubit singlet states 
during the QPC measurement. 
Two-qubit DM is analytically solved and perfectly analyzed.

%%%%%%%%%%%%%%%%%%%%%%%%%%%%%%%%%%%%%%%%%%%%%%%%%%%%%%%%%%%%%%
%-- {\it Formulation}---
%%%%%%%%%%%%%%%%%%%%%%%%
The Hamiltonian for the combined qubits and the QPC is written as
$H = H_{\rm qb}\!+\!H_{\rm qpc}\!+\!H_{\rm int}$.  $H_{\rm qb}$ describes the
interacting $N$ qubits (Fig.~\ref{QPC}) :
%%%%%%%%%%%%%%%%%%%%%%%%%%
$ %\begin{equation}
H_{\rm qb}$ $\!=\!\sum_{i=1}^N \!
\left(\Omega_i \sigma_{i x} \!+\!
\epsilon_i \sigma_{i z} \right)$
$\!+\! \!\sum_{i=1}^{N\!-\!1}
 \!J_{i,i+1} \sigma_{i z} \sigma_{i+1z}, 
$ %\end{equation}
%%%%%%%%%%%%%%%%%%%%%%%%%%
where $\Omega_i$ and $\epsilon_i$ are the inter-QD tunnel coupling and
energy difference (gate bias) within each qubit.  
Here %we use 
the spin operators are expressed by 
annihilation operators of an electron in the upper and
lower QDs of each qubit.  
$J_{i,i+1}$ is a coupling constant between two nearest qubits,
originating from capacitive couplings in the QD system \cite{Tanamoto}.
$|\!\uparrow\rangle$ and $|\!\downarrow\rangle$ refer
to the two single-qubit states in which the excess charge is localized in the
upper and lower dot, respectively.  
%
%%%%%%%%%%%%%%%%%%%%%%%%%%
$ %%\begin{equation}
H_{\rm qpc}=$ $\! \!\! \sum_{\alpha=L,R }\! \sum_{\ i_\alpha s} 
\!\! E_{i_\alpha} c_{i_\alpha s}^\dagger c_{i_\alpha s}$ 
$\!+\!\sum_{i_L,i_R s} V_{s} (c_{i_Ls}^\dagger c_{i_Rs} +
c_{i_Rs}^\dagger c_{i_Ls} )
$ %\end{equation}
%%%%%%%%%%%%%%%%%%%%%%%%%%%
describes the QPC.
Here $c_{i_{L}s}$($c_{i_{R}s}$) $s\!=\!\uparrow,\downarrow$
is the annihilation operator of an electron
in the $i_L$th ($i_R$th) level ($i_L(i_R)=1,...,n)$ of the left(right)
electrode.
$H_{\rm int}$ is the (capacitive) interaction between the qubits and the QPC,
that induces {\it dephasing} between different eigenstates of $\sigma_{iz}$
\cite{Zanardi}.
Most importantly, it contains the fact that localized charge near the QPC
increases the energy of the system electrostatically, thus affecting the
tunnel coupling between the left and right electrodes: 
%%%%%%%%%%%%%%%%%%%%%%%%%%
\begin{equation}
H_{\rm int} =  \sum_{i_L,i_R,s} 
\left[\sum_{i=1}^N \delta V_{is}\sigma_{i z}\right]
( c_{i_Ls}^\dagger c_{i_Rs} + c_{i_Rs}^\dagger c_{i_Ls} ) .
\end{equation}
%%%%%%%%%%%%%%%%%%%%%%%%%%
Note that the case where $\delta V_i$ is independent of the qubit 
corresponds to {\it collective environment}, 
thus the DF states are realized. 
%- The problem is the robustness of the DF states 
%- when $\delta V_i$ fluctuates much, 
%- which is the subject of this paper.
Hereafter we neglect the spin dependence of $V$ and $\delta V_i$.
We assume that the tunneling rate through the $N$ qubits,  $\Gamma$, 
is composed of direct series of 
each $N$ tunneling rate near $i$-th qubit, $\Gamma_{i}$, 
such as $\Gamma^{-1}=\sum_{i} \Gamma_{i}^{-1}$, 
where $\Gamma_{i}$ is defined as
$\Gamma_{i}^{(\pm)} \equiv 2\pi \wp_{L}\wp_{R} (v_d/N)
|V\pm \delta V_{i} |^2$ ($\wp_{L}$ and $\wp_{R}$ 
are the density of states of the electrodes 
at the Fermi surface) depending on the qubit state 
$\sigma_{iz}\!=\!\pm\! 1$.
The strength of measurement is parameterized by 
$\Delta \Gamma_{i}$ as 
$\Gamma_{i}^{(\pm)}\!=\!\Gamma_{i0}\!\pm\Delta \Gamma_{i}$.
We call $|\downarrow \downarrow \rangle$, $|\downarrow
\uparrow \rangle$, $|\uparrow \downarrow \rangle$, and $|\uparrow \uparrow
\rangle$ as $|A\rangle \sim |D\rangle$ respectively, 
and four-qubit states are written by $|AA\rangle$, $|AB\rangle$
...$|DD\rangle$. For uniform two qubits, 
$\Gamma_A\!=\!\Gamma_0(1\!-\!\zeta)/2$, 
$\Gamma_B\!=\!\Gamma_C$ $\!=\!\Gamma_0(1\!-\!\zeta^2)/2$ and
$\Gamma_D\!=\!\Gamma_0(1\!+\!\zeta)/2$ with $\zeta\equiv \Delta\Gamma/\Gamma_0$.
The DM equations of $N$ qubits and detector at zero temperature 
are derived similar to Ref.\cite{TanaHu} as
%%%%%%%%%%%%%%%%%%%%%%%%%%%%%%%%%%%%%%%%%%%%%%
%{DM equations}
%%%%%%%%%%%%%%%%%%%%%%%%%%%%%%%%%%%%%%%%%%%%%%
%\small
\begin{eqnarray}
%%%%%% 
\frac{d \rho_{z_1z_2}}{dt}&\!=\!&i[J_{z_2}\!-\!J_{z_1}\!]
\rho_{z_1z_2}
\!-\! i\sum_{j=1}^{N} \Omega_j (\rho_{g_j(z_1),z_2}\!-\!\rho_{z_1,g_j(z_2)})
\nonumber \\
&\!-\!&\left[ \sqrt{\Gamma^{z_1}}\!-\!\sqrt{\Gamma^{z_2}}\right]^2
\rho_{z_1z_2}
\label{eqn:dm}
\end{eqnarray}
\normalsize
where $z_1,z_2=AA,AB,...,DD$ for four qubits (256 equations) and 
$z_1,z_2=A,B,C,D$ for two qubits (16 equations). 
%----%
$J_{AA} \!=\! \sum_i^4 \epsilon_i\!+\!J_{12}\!+\!J_{23}$, 
$J_{AB} \!=\! \sum_i^3 \epsilon_i\!-\!\epsilon_4\!+\!J_{12}\!-\!J_{23}$, 
...,
%- $J_C \!=\!-\epsilon_L\!+\!\epsilon_R\!-\!J$, 
$J_{DD} \!=\!-\sum_i^4\epsilon_i\!+\!J_{12}\!+\!J_{23}$.
%- $J_A \!=\! \epsilon_L\!+\!\epsilon_R\!+\!J$, 
%- $J_B \!=\! \epsilon_L\!-\!\epsilon_R\!-\!J$, 
%- $J_C \!=\!-\epsilon_L\!+\!\epsilon_R\!-\!J$, 
%- $J_D \!=\!-\epsilon_L\!-\!\epsilon_R\!+\!J$.
%
$g_l(z_i)$ and $g_r(z_i)$ are introduced for the sake of notational 
convenience and determined by the relative positions between qubit
states. For two qubits, 
%%%%%%
$g_1(A)=B$, $g_2(A)=C$, 
$g_1(B)=A$, $g_2(B)=D$, 
$g_1(C)=D$, $g_2(C)=A$, 
$g_1(D)=C$, $g_2(D)=B$. 
Because the detector current is 
$I=e\sum_{z=A,..,D} \Gamma^z\rho_{zz}$, we have 
$\Delta T/T\sim\Delta \Gamma/\Gamma_0(=\zeta)$\cite{tomography}.
In the two-qubit case, we use entangled Bell basis:
$|a\rangle \!\equiv$
$(|A\rangle $+$|D\rangle )/\sqrt{2}$, 
$|b\rangle \!\equiv$
$(|A\rangle$ -$|D\rangle )/\sqrt{2}$, 
$|c\rangle \!\equiv$
$(|B\rangle$ +$|C\rangle )/\sqrt{2}$, 
$|d\rangle \!\equiv$
$(|B\rangle$ -$|C\rangle )/\sqrt{2}$.

{\it Decoherence rates}--
From Eq.~(\ref{eqn:dm}), the dephasing is expected 
to be relevant to the {\it dephasing rate}
${\it \Gamma}_{\rm d}(z_1,z_2)
\equiv[\sqrt{\Gamma_{z_1}}\!-\!\sqrt{\Gamma_{z_2}}]^2$.
To see the decoherence effect explicitly, we study 
time-dependent {\it fidelity}, 
$F(t)\equiv {\rm Tr}[\rho(0) \rho'(t)]$ 
on the rotating coordinate as 
$ 
\hat{\rho}'(t)=e^{i\sum\Omega_i'\sigma_{ix} t} \hat{\rho} (t) 
e^{-i\sum\Omega_i'\sigma_{ix} t}
$ ($\Omega_i'\equiv \sqrt{\Omega_i^2\!+\!\epsilon_i^2/4}$) to eliminate 
the bonding-antibonding coherent oscillations of free qubits
(trace is carried out over qubit states).
$F(t)$ can be expanded in time as 
$F(t)\!=\!1\!-\!\sum_{n\!=\!1} (1/n!)(t/\tau^{(n)})^n$
where $1/\tau^{(n)} =\{-{\rm Tr}[\rho(0)d^n\rho(0)/dt^n]\}^{1/n}$ 
({\it decoherence rates}). 
Using Eq.(\ref{eqn:dm}) for two qubits, 
$1/\tau^{(1)}_c\!=\!1/\tau^{(1)}_d\!=\!(1/2){\it \Gamma}_{\rm d}(B,C)$
$\!\sim\! (\Gamma_0/8)(1\!-\!\zeta^2)\zeta^2\eta^2$
when $|B\rangle$ and $|C\rangle$ are not symmetric such as 
the left qubit has a local fluctuation 
$\Gamma^{(\mp)}(1\!-\!\eta)$.
Moreover, 
$(1/\tau^{(2)}_c)^2\!=\!(\Omega_1\!+\!\Omega_2)^2
\!+\!(\epsilon_1\!-\!\epsilon_2)^2\!+\!{\it \Gamma}_{\rm d}^2(B,C)/4$ and 
$(1/\tau^{(2)}_d)^2\!=\!(\Omega_1\!-\!\Omega_2)^2
\!+\!(\epsilon_1\!-\!\epsilon_2)^2\!+\!{\it \Gamma}_{\rm d}^2(B,C)/4$.
Thus,
the symmetry-breaking terms start from $O(\eta^2)$\cite{Lidar}. 
%--Note that $|c\rangle$ has 
%-- the same lowest decoherence rate as the singlet state. 
We have similar expressions for three-qubit DF {\it subsystem} 
and four-qubit DF {\it subspace}.
For $|\Psi^{[3]}_1\rangle\equiv (|010\rangle\!-\!|100\rangle)/\sqrt{2}$
\cite{Knill}, 
$1/\tau^{(1)}_{1[3]}\!=\!(1/2){\it \Gamma}_{\rm d}(010,100)$ and for 
$|\Psi_1^{[4]}\rangle_{(1234)}$,
%###############
\begin{equation}
\frac{1}{\tau^{(1)}_{1[4]}}=\frac{1}{8}\sum_{z_1,z_2=\tiny BB,BC,CB,CC}
{\it \Gamma}_{\rm d}(z_1,z_2),
\end{equation}
both of which also start from $O(\eta^2)$
(decoherence rate of $|\Psi_2^{[4]}\rangle_{(1234)}$ has a similar but more complicated form). 
The robustness of $N\ge\!3$ qubits
changes depends on which qubit includes local fluctuation. 
For example, if the leftmost qubit of $|\Psi^{[3]}_1\rangle$ fluctuates, 
$1/\tau^{(1)}_{1[3]}\sim 
(\Gamma_0/(3\!+\!\zeta)^3)(1\!-\!\zeta^2)\zeta^2\eta^2$, 
but $1/\tau^{(1)}_{1[3]}=0$ when the rightmost qubit fluctuates.
In the latter case, there is a symmetry between the leftmost and 
middle qubit and we can say that if some symmetry remains, DF states are robust.
To see the dependence of spatial arrangement of qubits, 
we also consider $|\Psi_3^{[4]}\rangle_{(1234)}$$\equiv$ 
$|\Psi_1^{[4]}\rangle_{(1423)}$ (hereafter we omit the subscript). 
$|\Psi_3^{[4]}\rangle$ is expected to be more susceptible than 
$|\Psi_1^{[4]}\rangle$, because the former is not a product of 
two singlet states as $|\Psi_1^{[4]}\rangle$ and  
less symmetric exchanges of qubit states are possible 
when qubit parameters fluctuate.  Thus, 
the DF states are strongly affected by
the distribution of local non-uniformities.

\begin{figure}
\begin{center}
\includegraphics[width=6.0cm]{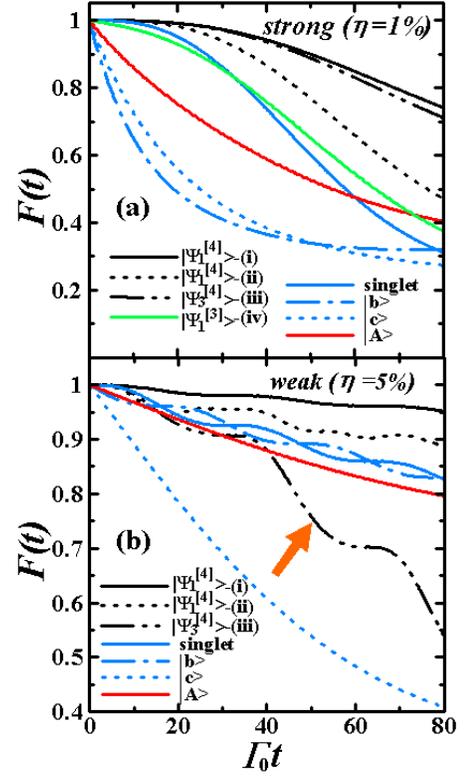}
\end{center}
\caption{Time-dependent fidelity 
of four-qubit DF states ($|\Psi_1^{[4]}\rangle$ and 
$|\Psi_3^{[4]}\rangle$)
%--($|\Psi_1^{[4]}\rangle$-(i), 
%--$|\Psi_3^{[4]}\rangle$-(ii), 
%--$|\Psi_3^{[4]}\rangle$-(iii)), 
three-qubit DF state of $|\Psi_1^{[3]}\rangle$, and 
two-qubit states (singlet, $|b\rangle$, $|c\rangle$, 
and product state $|A\rangle$) under various fluctuations : 
(i)$\Omega_3\!=\!(1\!-\!\eta)\Omega$,
$\epsilon_3\!=\!\eta\Gamma_0$ and 
$\Gamma_3^{(\pm)}\!=\!(1\!-\!\eta)\Gamma^{(\pm)}$. 
(ii)$\Omega_2\!=\!\Omega_3\!=\!(1\!-\!\eta)\Omega$,
$\epsilon_2\!=\!\epsilon_3\!=\!\eta\Gamma_0$ and 
$\Gamma_2^{(\pm)}=\Gamma_3^{(\pm)}\!=\!(1\!-\!\eta)\Gamma^{(\pm)}$.
(iii)$\Omega_4\!=\!(1\!-\!\eta)\Omega$,
$\epsilon_4\!=\!\eta\Gamma_0$ and 
$\Gamma_4^{(\pm)}\!=\!(1\!-\!\eta)\Gamma^{(\pm)}$.
(iv)$\Omega_2\!=\!(1\!-\!\eta)\Omega$,
$\epsilon_2\!=\!\eta\Gamma_0$ and 
$\Gamma_2^{(\pm)}\!=\!(1\!-\!\eta)\Gamma^{(\pm)}$.
(iv) is also applied to the two qubits. 
%-
(a) $\eta$=0.01 and $\zeta$=0.6 (strong measurement), 
(b) $\eta$=0.05 and $\zeta$=0.2 (weak measurement). 
$\Omega=2\Gamma_0$, $J_{ij}=0$ $\epsilon_i=0$.
}
\label{FID}
\end{figure}

\begin{figure}
\begin{center}
\includegraphics[width=4.0cm]{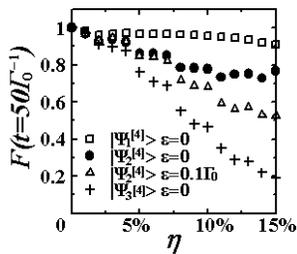}
\end{center}
\caption{Fidelity of four-qubit DF states 
at $t\!=\!50\Gamma_0^{-1}$in the case of (iii) 
as a function of non-uniformity.
$\Omega=2\Gamma_0$, $J_{ij}=0$ and $\zeta$=0.2.
}
\label{eta}
\end{figure}

%%%%%%%%%%%%%%%%%%%%%%%%%%%%%%%%%%%%%%%%%%%%%%%%%%%%%%%%%%%%%%
%%%%%%%%%%%%%%%%%%%%%%%%%%%%%%%%%%%%%%%%%%%%%%%%%%%%%%%%%%%%%%
%%{\it Numerical calculations}--
Numerical calculations support these analyses. 
Here we add fluctuations {\it locally} to $\Omega_i$, 
$\epsilon_i$ and $\Gamma_i$ respectively to 
various DF states of $N\stackrel{<}{=}4$ qubits.
In case (i), 
only 3rd qubit fluctuates 
as $\Omega_3\!\rightarrow\! \Omega_3(1\!-\!\eta)$, 
$\epsilon_3\!\rightarrow\! \epsilon_3(1\!-\!\eta)$ and
$\Gamma_3\!\rightarrow\! \Gamma_3(1\!-\!\eta)$. 
In case (ii), the 2nd and 3rd qubits fluctuate. In case (iii)
only 4th qubit fluctuates.
Fig. \ref{FID} (a) shows a time-dependent $F(t)$ 
in a strong measurement case of $\zeta=0.6$ of 1\% 
fluctuations ($\eta\!=\!0.01$) 
and Fig.\ref{FID} (b) shows that of a weak measurement case of 
$\zeta=0.2$ of 5\% fluctuations ($\eta\!=\!0.05$), both in degeneracy point 
$\epsilon=0$ ({\it relaxation} decoherence region\cite{Makhlin,You}). 
Although Fig. \ref{FID} (a) indicates that the four-qubit DF states are fairly 
robust, Fig. \ref{FID} (b) shows that the large fluctuation (5\%) greatly 
degrades  $|\Psi_3^{[4]}\rangle$ in case (iii).
If we take $\Gamma_0$=100MHz, its lifetime ($F(t)>0.75$) is about
$50\Gamma_0^{-1}\sim 500n$s, which is much shorter than 
lowest order estimation of $\eta^{-2}\Gamma_0^{-1}\sim 40\mu$s. This 
shows that higher symmetry-breaking perturbation terms cannot be 
neglected and there is a case where a product state is more preferable
than four-qubit DF states.
$F(t)$ for $|\Psi_2^{[4]}\rangle$ 
is in the same order of $|\Psi_1^{[4]}\rangle$, which means that 
DF states composed of many entangled states seem fairly robust 
at the degeneracy point. 
The three-qubit DF states are fairly robust for larger $\eta$,
and show susceptibilities to measurement strength 
as the singlet state and other Bell states (Fig.\ref{FID}(a)).
In the case of finite bias $\epsilon$ 
({\it pure dephasing} region), the effect of which 
appears from $1/\tau^{(2)}$, 
$F(t)$ of $|\Psi_2^{[4]}\rangle$ and $|\Psi_3^{[4]}\rangle$
seem more susceptible than the singlet state and the singlet 
type $|\Psi_1^{[4]}\rangle$ (Fig.\ref{eta}).
The other Bell states and three-qubit DF states are less 
susceptible than $|\Psi_2^{[4]}\rangle$ and $|\Psi_3^{[4]}\rangle$, 
which would again be due to the higher order symmetry-breaking terms.
To summarize these results, 
DF states of many qubit ($N\!=\!4$) are robust 
for most cases
%- to small fluctuations ($\sim$ 1\%) , near the degeneracy point,
but there is a case where even product states 
might be better under large symmetry-breaking 
fluctuations ($\eta \stackrel{>}{\sim}5$\% )
%-and finite $\epsilon$ 
in the long term. 
%%%%%%%%%%%%%%%%%%%%%%%%%%%%%%%%%%%%%%%%%%%%%%%%%%%%%%%%%%%%%%%%%
%%%%%%%%%%%%% Fig.2
%%%%%%%%%%%%%%%%%%%%%%%%%%%%%%%%%%%%%%%%%%%%%%%%%%%%%%%%%%%%%%%%%
%%%%%%%%%%%%%%%%%%%%%%%%%%%%%%%%%%%%%%%%%%%%%%%%%%%%%%%%%%%%%%%%%
%%%%%%%%%%%%% Fig.3
%%%%%%%%%%%%%%%%%%%%%%%%%%%%%%%%%%%%%%%%%%%%%%%%%%%%%%%%%%%%%%%%%
%%%%%%%%%%%%%%%%%%%%%%%%%%%%%%%%%%%%%%%%%%%%%%%%%%%%%%%%%%%%%%%%%
%%%%%%%%%%%%% Fig.4
\begin{figure}
\begin{center}
\includegraphics[width=4.0cm]{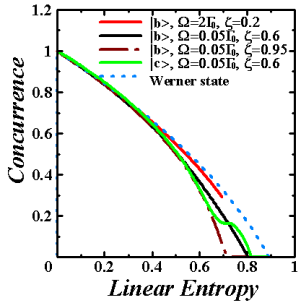}
\end{center}
\caption{Linear entropy $S_L$ and concurrence $C$ plane
under the measurement during $0\stackrel{<}{=} t \stackrel{<}{=} 100\Gamma_0$
for two qubit states $|b\rangle$ and $|c\rangle$ ($\epsilon=0$). 
In weak measurement case of 
$\zeta\stackrel{<}{=} 0.2$, 
%- $\Delta \Gamma\stackrel{<}{=} 0.2\Gamma_0$, 
$C$ does not degrade to 0.
All data start closely to Werner state.}
\label{purity_plane}
\end{figure}
%%%%%%%%%%%%%%%%%%%%%%%%%%%%%%%%%%%%%%%%%%%%%%%%%%%%%%%%%%%%%%%%%

%%%%%%%%%%%%%%%%%%%%%%%%%%%%%%%%%%%%%%%%%%%%%%%%
%%%%%%%%%%%%%%%%%%%%%%%%%%%%%%%%%%%%%%%%%%%%%%%%
{\it Analytical solution for two-qubit case}--
The probability that the unexpected non-uniformities
induce symmetry-breaking effects 
becomes higher as the number of qubits increases as shown above. 
Because preparing many solid-state qubits is not easy, the redundancy of 
coding qubits is a trade-off against the fabrication difficulty. Thus, 
using the singlet state and one of the non-DF states,
$\{ \rho_{aa}, \rho_{bb}, \rho_{cc}\}$ in two qubit space 
would sometimes be a realistic solution to construct two logical states 
$|0\rangle_L$ and $|1\rangle_L$. Here we investigate 
which of $\{ \rho_{aa}, \rho_{bb}, \rho_{cc}\}$ is appropriate 
for the second basis in two-qubit space. 
When we move to four Bell bases
%#####
under the conditions that the two qubits are identical with 
no interaction between them ($J_{ij}\!=\!0$) and no bias 
$\epsilon_i\!=\!0$ (collective environment), Eqs.(\ref{eqn:dm}) 
are divided into the following five groups:
%#############################
\begin{eqnarray}
% {dd}
& & \ \ \ \ \dot{\rho}_{dd}= 0
\\
%% {bd}
& & \ \ \ \ \dot{\rho}_{bd}=
\!-\!\gamma^{B} \rho_{bd}
\\
%%---------------------------%%
& & \left\{\begin{array}{ll}
%% {ad}
\dot{\rho}_{ad}&=-2i\Omega \rho_{cd}
\!-\!\gamma^{B} \rho_{ad}
\\
% {cd}
\dot{\rho}_{cd}&=-2i\Omega \rho_{ad}
\end{array}
\right. \\
%%---------------------------%%
& & \left
\{\begin{array}{ll}
% {ab}
\dot{\rho}_{ab}&=-2i\Omega \rho_{cb}
\!+\!\frac{1}{2}\gamma^{D} (\rho_{ba}-\rho_{ab})
\\
%% {bc}
\dot{\rho}_{bc}&=2i\Omega \rho_{ba}
\!-\!\gamma^{B} \rho_{bc}
\end{array}
\right.  \\
%%---------------------------%%
% {aa}
& & \left\{\begin{array}{ll}
\dot{\rho}_{aa}&=2i\Omega (\rho_{ac}-\rho_{ca})
\!-\!\frac{1}{2}\gamma^{D} (\rho_{aa}-\rho_{bb})
\label{eqn:aa} \\
% {bb}
\dot{\rho}_{bb}&=
\frac{1}{2}\gamma^{D} (\rho_{aa}-\rho_{bb})
\label{eqn:bb} \\
% {cc}
\dot{\rho}_{cc}&=-2i\Omega (\rho_{ac}-\rho_{ca})
\label{eqn:cc} \\
%% {ac}
\dot{\rho}_{ac}&=2i\Omega (\rho_{aa}-\rho_{cc})
\!-\!\gamma^{B} \rho_{ac}
\label{eqn:ac} 
\end{array}
\right.\label{eqn:dm2}
\end{eqnarray}
where 
%- $\gamma^{D}\!=\!(\sqrt{\Gamma_A}\!-\!\sqrt{\Gamma_D})^2$ and 
%- $\gamma^{B}$=$ (\sqrt{\Gamma_A}$-$\sqrt{\Gamma_B})^2$.
$\gamma^{D}\!=\!{\it \Gamma}_{\rm d}(A,D)$ and 
$\gamma^{B}$=$ {\it \Gamma}_{\rm d}(A,B)$.
These equations can be solved analytically. 
First, singlet state $\rho_{dd}$ is time-independent (DF state) and 
$\rho_{bd}(t)=\rho_{bd} (0)e^{-\gamma^B t}$. 
$\rho_{ad}(t)$ and $\rho_{cd}(t)$ behave like 
$e^{-(\gamma^B\pm\sqrt{(\gamma^{B})^2-16\Omega^2})t}$, and 
$\rho_{ab}(t)$ and $\rho_{bc}(t)$ behave like 
$e^{-(\gamma^B\!+\!\gamma^D
\!\pm\!\sqrt{(\gamma^B\!+\!\gamma^D)^2-16\Omega^2})t/2}$,
thus decay rates of $\rho_{ad}(t)$, $\rho_{cd}(t)$, 
$\rho_{ab}(t)$,and $\rho_{bc}(t)$ depend on whether
$4\Omega > \gamma^B$, $\gamma^B+\gamma^D$ or 
$4\Omega < \gamma^B$, $\gamma^B+\gamma^D$, respectively. 
From Eq.(\ref{eqn:dm2}),
%###################
%- \begin{eqnarray}
$\rho_{aa}(t)\!+\!\rho_{bb}(t)\!+\!$
$\rho_{cc}(t)$ is conserved and 
%=$\rho_{aa}(0)+\rho_{bb}(0)+\rho_{cc}(0)$.
$\rho_{ac}(t)\!+\!\rho_{ca}(t) 
=\! (\rho_{ac}(0)+\rho_{ca}(0))e^{-\gamma_B t}$.
%--%
%-\end{eqnarray}
Thus, $\rho_{aa}(t)$, $\rho_{bb}(t)$, $\rho_{cc}(t)$ and 
$\rho_{ac}(t)$ can be analytically obtained by solving
three-order polynomial equations. 
In the case of high qubit oscillation 
$\Omega \gg \gamma^B,\gamma^D$ and weak measurement 
such as $\zeta \ll 1$, 
we obtain $\gamma^{D} \sim 4 \gamma^{B}$ 
in order of $\zeta^2$.
%-- due to relation $\Gamma=(\Gamma_1 \Gamma_2)/(\Gamma_1\!+\!\Gamma_2)$. 
Then we have the eigenvalues of time-dependent 
matrix equations for  $\{\rho_{aa},\rho_{cc}$, 
$\rho_{ac}-\rho_{ca}\}$
in order of $\delta \equiv \gamma^{B}/(4\Omega)$ as
%----%
$-3\gamma^B$, 
$-\gamma^B\!\pm\!4\Omega i\left(1\!-\!5/3 \delta^2\right)$,
which shows that the period of qubit oscillation is 
delayed by the measurement through $\gamma^B$ and $\gamma^D$. 
Thus we found that the qubit behavior strongly depends 
on the  relative magnitude of $\Omega$ to $\gamma$ 
and this is also an important factor in the selection 
of the possible candidate of the logical basis. 
In Fig.~\ref{FID}(b) of $\Omega=2\Gamma>\gamma^D\sim0.2\Gamma$, 
$F(t)$s of $|b\rangle $ and $|d\rangle$ 
are larger than those of $|c\rangle$ (and $|a\rangle $). 
In Ref.\cite{QPC}, we argued the coherent oscillation 
of the DM without the detector and showed that 
the DM for $|b\rangle $ and $|d\rangle$ are time-independent 
at $\epsilon\!=\!0$. This indicates that $|b\rangle $ 
is the candidate in the $\Omega\gg\gamma_B,\gamma_D$ region, reflecting that 
$|b\rangle $  becomes an unpolarized triplet suffering 
 less degradation from the repulsive Coulomb interaction 
of the detector. On the other hand, in the $\Omega=0$ case 
which can be in the limit of 
$\Omega \ll \gamma^D, \gamma^B$, 
Eq.~(\ref{eqn:dm}) is easily solved even with a finite bias $\epsilon$
and $\rho_{cc}$ is found to be a time-independent 
unpolarized triplet, being the candidate of logical state.

Next, we compare   $\rho_{bb}$ and $\rho_{cc}$ 
in the  {\it purity plane}, that is a relation 
between {\it linear entropy} and {\it concurrence}\cite{Munro}.
Linear entropy ${\sl S}_L=4/3(1-{\rm Tr} (\hat{\rho}^2))$
expresses purity of qubits,
ranging from 0 (pure state) and 1 (maximally-mixed state).
From Eq.(\ref{eqn:dm2}), 
$S_L=4/3[1-\rho_{aa}^2-\rho_{bb}^2-\rho_{cc}^2
-2|\rho_{ac}|^2 ]$, and we have
%%%%%%%%%%%%%%%
%%--\begin{equation}
%% \frac{d S_L}{dt}=\frac{2}{3}\gamma^D (\rho_{bb}-\rho_{aa})^2
%% +\frac{8}{3}\gamma^B|\rho_{ac}|^2  
%%\end{equation}
$d S_L/dt\!=\!(2/3)\gamma^D (\rho_{bb}\!-\!\rho_{aa})^2$
$+\!(8/3)\gamma^B|\rho_{ac}|^2$. 
Concurrence ${\sl C}$, which is a measure of entanglement, 
is also given from Eq.(\ref{eqn:dm2}).
Starting from $\rho_{bb}(0)\!=\!1$, 
at $t\!\sim\!0$ in the case of $\Omega \!\gg\! \gamma_D\!=\!4\gamma_B$,
we have $C\sim(-1\!+\!4e^{-3\gamma^Bt})/3$ (
$
d \sl C/dt\!\sim\! -\gamma^D\rho_{bb}/2
$)
and $S_L\sim 8/9(1\!-\!e^{-6\gamma^Bt})$. 
Then $dC/dS_L\sim $-3/4 near $t\sim0$. 
%-- $C$ and $S_L$ are determined mainly 
%-- by the dephasing rates and the ratio is constant 
If we start $\rho_{cc}(0)\!=\!1$ in the case of 
$\Omega \!\ll\!\gamma_D\!=\!4\gamma_B$, 
we have $C\sim (-1\!+\!4e^{-(2/3)\Omega t})$ and 
$S_L\sim 8/9 (1\!-\!e^{-(4/3)\Omega t})$, and thus 
$dC/dS_L\sim $-3/4.
%%%%%%%%%%%%%%%%%%%%%%%%%%%%
If we check the Werner state
$\gamma_w |b\rangle \langle b|\!+\!(1\!-\!\gamma_w)/4
\hat{I}_2\otimes \hat{I}_2$ where $\hat{I}_2$ is 2x2 unit matrix
($1>\gamma_w>0$)
\cite{Werner,Munro}, 
%- $S_L^{(w)}=1-\gamma_w^2$ and
%- $C^{(w)}=(3\gamma_w-1)/2$, and 
we obtain $dC^{(w)}/dS_L^{(w)}=-3/4$ 
at $\gamma_w=1$.
Thus both $\rho_{bb}$ and $\rho_{cc}$ coincide 
with the Werner state at $t\sim0$, which shows that the two states
are good entangled states, because the Werner state 
is a mixture of the maximally entangled state\cite{Munro}. 
Figure~\ref{purity_plane} shows that both $\rho_{bb}$ and $\rho_{cc}$ 
evolve close to the Werner state. 
%- , and deviates slightly from Werner state as time goes by. 
%%%%@check
%In the figure, the result of $\Omega=\Gamma$ and
%$\Omega=2\Gamma$ are almost same if $\Omega >\gamma^B,\gamma^D$. 
Thus, the two states behave similarly in the purity plane, 
and are equal candidates.

%%%%%%%%%%%%%%%%%%%%%%%%%%%%%%%%%%%%%%%%%%%%%%%%%%%%%%
%-- {\it Discussion ---} 
The noise spectrum $S(\omega)$ of the QPC without qubits 
is given by $S(\omega)=e^2\Gamma/\pi$ (white noise) 
in the present model, thus, the shot noise affects qubit states 
in full frequency domain.
Astafiev {\it et al.}\cite{Astafiev} experimentally showed
that the main causes of the noise in the Josephson qubits
are $f$ noise and the background charge noises or $1/f$ noise, 
which we do not include. These noises would locally affect 
qubits and degrade the robustness of the DF states 
more than discussed here.
%- Although there are theories for phonons\cite{Leggett} 
%- and $1/f$ noise\cite{Makhlin}, their extension to more than two qubits 
%- is not straightforward, due to the difficulty 
%- in the treatment of many-body interaction
%- between the large degrees of freedom of environments
%- and spatially separated qubits. 
%- We showed only one of the ideal cases, which will be 
%- detected {\it eg.} at very low temperature 
%- where the effect of phonons or other decoherence 
%- origins are reduced. 

%%%%%%%%%%%%%%%%%%%%%%%%%%%%%%%%%%%%%%%%%%%%%%%%%%%%%%
%{\it Conclusions}--
In conclusion,
we have solved master equations of many
qubits and QPC detector, and discuss the robustness of 
DF states under the large non-uniformities ($\sim$ 5\%)
during the long time period.
Two-qubit non-DF states are shown to be one solution
in constructing logical qubits.

We thank N. Fukushima, X. Hu, M. Ueda, T. Fujisawa 
S. Ishizaka and K. Uchida for fruitful discussions.

%%%%%%%%%%%%%%%%%%%%%%%%%%%%%%%%%%%%%%%%%%%%%%

%--\clearpage

\begin{thebibliography}{99}
\bibitem{Chuang}
M. A. Nielsen and I. Chuang, {\it Quantum Computation and 
Quantum Information}, (Cambridge Univ. Press, 2000).

\bibitem{Zanardi}
P. Zanardi and M. Rasetti, Phys. Rev. Lett. {\bf 79}, 3306(1997).

\bibitem{Lidar}
D. A. Lidar {\it et al.}, %-- D. Bacon, and K. B. Whaley, 
Phys. Rev. Lett. {\bf 82}, 4556 (1999).
D. Bacon {\it et al.}, %-- D. A. Lidar and K. B. Whaley,
Phys. Rev. A {\bf 60}, 1944 (1999).
% "Concatenating Decoherence-Free Subspaces with Quantum Error Correcting Codes
% An operator sum representation is derived for a decoherence-free subspace 
% (DFS) and used to (i) show that DFSfs are the class of quantum error
% correcting codes (QECCfs) with fixed, unitary recovery operators and 
% (ii) find explicit representations for the Kraus operators of collective
% decoherence. We demonstrate how this can be used to construct 
% a concatenated DFS-QECC code, which protects against collective decoherence
% perturbed by independent decoherence. The code yields an error threshold 
% which depends only on the perturbing independent decoherence rate.

\bibitem{Knill}
E. Knill {\it et al.}, %-- R. Laflamme and L. Viola, 
Phys. Rev. Lett. {\bf 84} 2525 (2000).
J. Kempe {\it et al.}, %-- D. Bacon, D. A. Lidar and K. B. Whaley, 
Phys. Rev. A {\bf 63}, 042307 (2001).

\bibitem{Bourennane}
M. Bourennane {\it et al.}, %--M. Eibl, S. Gaertner, C. Kurtsiefer,
%-- A. Cabello, and H. Weinfurter, 
Phys. Rev. Lett. {\bf 92}, 107901 (2004).
% " Decoherence-Free Quantum Information Processing with Four-Photon 
% Entangled States"
% Decoherence-free states protect quantum information from collective 
% noise, the predominant cause of decoherence in current implementations 
% of quantum communication and computation. Here we demonstrate that 
% spontaneous parametric down conversion can be used to generate 
% four-photon states which enable the encoding of one qubit in a 
% decoherence-free subspace. The immunity against noise is verified 
% by quantum state tomography of the encoded qubit. We show that 
% particular states of the encoded qubit can be distinguished by local
% measurements on the four photons only.

\bibitem{Altepeter}
J. B. Altepeter {\it et al.}, 
%-- P.G. Hadley, S.M.Wendelken, A. J. Berglund, and P.G. Kwiat,
Phys. Rev. Lett. {\bf 92}, 147901 (2004).
% " Experimental Investigation of a Two-Qubit Decoherence-Free Subspace"
% 1Department of Physics, University of Illinois at Urbana-Champaign, 
% Urbana, % Illinois 61801-3080, USA
% 2Physics Division, P-23, Los Alamos National Laboratories, Los Alamos, 
% New %Mexico 87545, USA
% We thoroughly explore the phenomenon of a decoherence-free subspace (DFS) 
% for two-qubit systems. Specifically, we both collectively and 
% noncollectively decohere entangled polarization-encoded twoqubit
% states using thick birefringent crystals. These results characterize 
% the basis-dependent effect of decoherence on the four Bell states, 
% the robustness of the DFS state against perturbations in the assumption 
% of collective decoherence, and the existence of a DFS for each type of 
% stable noncollective decoherence. Finally, we investigate the effects 
% of collective and noncollective dissipation.

\bibitem{Viola}
L. Viola {\it et al.},
%--  E.M. Fortunato, M.A. Pravia, E. Knill, R. Laflamme, D.G. Cory.
Science, {\bf 293}, 2059 (2001).
% "Exper-imental Realization of Noiseless Subsystems for Quantum 
% Information Processing".

\bibitem{Nakamura}
Y. Nakamura {\it et al.}, 
%--Yu.A. Pashkin, and J.S. Tsai, 
Nature {\bf 398}, 786 (1999);
% Yu. A. Pashkin, T. Yamamoto, O. Astafiev, Y. Nakamura,
% D.V. Averin, and J.S. Tsai, Nature {\bf 421}, 823 (2003);
T. Yamamoto {\it et al.}, 
%-- Yu. A. Pashkin, O. Astafiev, Y. Nakamura and J. S. Tsai, 
{\it ibid.} {\bf 425}, 941 (2003). 

\bibitem{Hayashi}
% T. Fujisawa, D. G. Austing, Y. Tokura, Y. Hirayama and S. Tarucha, 
% Nature {\bf 419}, 278 (2002); Phys. Rev. Lett. {\bf
% 88}, 236802 (2002);
T. Hayashi {\it et al.}, 
%-- T. Fujisawa, H. D. Cheong, Y. H. Jeong, and Y. Hirayama 
Phys. Rev. Lett. {\bf 91}, 226804 (2003).

\bibitem{Makhlin}
%Y. Makhlin, G. Sch\"{o}n, and A. Shnirman, 
% Rev. Mod. Phys. {\bf 73}, 357 (2001).
Y. Makhlin, G. Sch\"{o}n, and A. Shnirman, 
Phys. Rev. Lett. {\bf 92} 178301 (2004).

\bibitem{You}
J. Q. You, X. Hu and F. Nori, cond-mat/0407423.

\bibitem{Uchida}
K. Uchida and S. Takagi, Appl. Phys. Lett. {\bf 82} 2916 (2003);
% "Carrier scattering induced by thickness fluctuation of 
% silicon-on-insulator film in ultrathin-body metal-oxide-semiconductor
% field-effect transistors"
% Advanced LSI Technology Laboratory, Toshiba Corporation 
% We demonstrate that carrier scattering induced by the thickness 
% fluctuation of a silicon-on-insulator ~SOI! film reduces electron 
% mobility in ultrathin-body metal-oxide-semiconductor field-effect
% transistors with SOI thickness, TSOI , of less than 4 nm at room 
% temperature and is the dominant scattering mechanism at low 
% temperatures. The thickness fluctuation of a nanoscaled SOI film
% induces large potential variations due to the difference of 
% quantum-confinement effects from one part to another, and thus 
% carrier scattering potentials are formed in the channel. It is shown that
% experimental electron mobility follows the theoretical TSOI dependence 
% and the expected temperature dependence of the scattering induced by 
% SOI thickness fluctuation. 
% H. Sakaki, T. Noda, K. Hirakawa, M. Tanaka, and T. Matsusue, Appl.
% Phys. Lett. {\bf 51}, 1934  (1987).

\bibitem{Landauer}
R. Landauer, Nature {\bf 272}, 1914 (1996).

\bibitem{Tanamoto}
T. Tanamoto, Phys. Rev. A {\bf 64}, 062306 (2001); {\it ibid.} {\bf 61},
022305 (2000).

\bibitem{Field}
M. Field {\it et al.}, %--C. G. Smith, M. Pepper, D. A. Ritchie, 
%-- J. E. F. Frost, G. A. C. Jones, and D. G. Hasko, 
Phys. Rev. Lett. {\bf 70}, 1311 (1993),
% "Measurements of Coulomb blockade with a noninvasive voltage probe"
% Cavendish Laboratory, Madingley Road, Cambridge CB3 OHE, United Kingdom
% We have investigated the behavior of a laterally confined QD in 
% close proximity to a one-dimensional channel in a separate electrical circuit.
% When this channel is biased in the tunneling regime the resistance is very
% sensitive to electric fields, and therefore is sensitive to the potential
% variations on the dot when it is showing Coulomb blockade oscillations. 
% This effect can be calibrated directly, allowing the Coulomb charging energy
% to be measured. We also found the activation energy of transport through the
% dot is much lower than expected.
%- \bibitem{Gardelis}
S. Gardelis {\it et al.}, %-- C. G. Smith, J. Cooper, D. A. Ritchie, 
%-- E. H. Linfield, Y. Jin,and M. Pepper, 
Phys. Rev. B {\bf 67}, 073302 (2003). 

\bibitem{Vandersypen}
L. M. Vandersypen {\it et al.}, 
%-- J. M. Elzerman, R. N. Schouten, L. H. Willems van Beveren,
%-- R. Hanson and L. P. Kouwenhoven,
Appl. Phys. Lett. {\bf 85}, 4394 (2004).

\bibitem{TanaHu}
T. Tanamoto and X. Hu, Phys. Rev. B {\bf 69} 115301 (2004).

\bibitem{tomography}
To differentiate the DF states, local measurements are 
required\cite{Bourennane}. In Fig.1, 
this means that other independent detectors are required or 
$\Gamma_i$ should be different for each qubit as in Ref.\cite{TanaHu}. 
However, because the aim of this paper is to see the 
effect of local non-uniformity, we take the simple setup 
of Fig.1.

\bibitem{QPC}
T. Tanamoto and X. Hu, cond-mat/0310293.

% \bibitem{Wiel}
% W. G. van der Wiel, S. De Franceschi, J. M. Elzerman, 
% T. Fujisawa, S. Tarucha and L. P. Kouwenhoven, 
% Rev. Mod. Phys. {\bf 75}, 1 (2003). %;
% S. Tarucha, D. G. Austing, T. Honda, 
% R. J. van der Hage and L. P. Kouwenhoven, 
% Phys. Rev. Lett. {\bf 77}, 3613 (1996).


\bibitem{Munro}
W. J. Munro {\it et al.}, %-- D. F. V. James, A. G. White, and P. G. Kwiat,
Phys. Rev. A {\bf 64}, 030302 (2001).
% "Maximizing the entanglement of two mixed qubits"
% Two-qubit states occupy a large and relatively unexplored Hilbert 
% space. Such states can be succinctly characterized by their degree 
% of entanglement and purity. In this article we investigate entangled 
% mixed states and present a class of states that have the maximum 
% amount of entanglement for a given linear entropy. 

\bibitem{Werner}
R. F. Werner, Phys. Rev. A {\bf 40}, 4277 (1989).
%" Quantum states with Einstein-Podolsky-Rosen correlations 
% admitting a hidden-variable model" /Reinhard F. Werner 
% A state of a composite quantum system is called classically correlated if 
% it can be approximated by convex combinations of product states, and 
% Einstein-Podolsky-Rosen correlated otherwise. Any classically correlated 
% state can be modeled by a hidden-variable theory and hence satisfies all
% generalized Bell's inequalities. It is shown by an explicit example 
% that the converse of this statement is false.  pp4277-4281

\bibitem{Astafiev}
O. Astafiev {\it et al.}, 
%Yu. A. Pashkin,1, Y. Nakamura,T. Yamamoto, and J. S. Tsai
cond-mat/0411216. 

%- \bibitem{Leggett}
%- A. J. Leggett, S. Chakravarty, A. T. Dorsey, M. P. A. Fisher, 
%- A. Garg and W. Zwerger, Rev. Mod. Phys. {\bf 59}, 1 (1987).
%- \bibitem{Gurvitz}
%- S. A. Gurvitz, Phys. Rev. B {\bf 56}, 15215 (1997).
%- \bibitem{Gurvitz}
%- S. A. Gurvitz and Ya. S. Prager, 
%- Phys. Rev. B {\bf 53}, 15932 (1996).
%- \bibitem{Wootters}
%- W. K. Wootters, Phys. Rev. Lett. {\bf 80}, 2245 (1998).
%- \bibitem{Schoelkopf}
%- R. J. Schoelkopf, P. Wahlgren, A. A. Kozhevnikov, P. Delsing,
%- and D. E. Prober, Science {\bf 280}, 1238 (1998).
%- \bibitem{Cain}
%- P.A. Cain, H. Ahmed, and D.A. Williams, J. Appl. Phys. {\bf 92}, 346 (2002);
%- Appl. Phys. Lett. {\bf 78} 3624 (2001).
%- \bibitem{Potok}
%- R.M. Potok, J. A. Folk, C.M. Marcus, V. Umansky, M. Hanson, and A. C. Gossard
%- Phys. Rev. Lett. {\bf 91}, 016802 (2003).

\end{thebibliography}
\end{document}